\documentclass[12pt,a4]{article}
\usepackage{bm}
\usepackage{graphics}
\usepackage{rotating}
\usepackage{epsfig}

\begin{document}
\begin{center}
{\Large \bf Weak decays of the $\bm{B_c}$ meson \footnote{Invited talk
    at the XVII International Workshop on High Energy Physics and
    Quantum Field Theory (QFTHEP 2003), Samara--Saratov, 4--11
    September 2003. }}\\
\vspace{4 mm}

D. Ebert$^a$, R. N. Faustov$^{a,b}$, V. O. Galkin$^{a,b}$

\vspace{4 mm}

${}^a$ Institut f\"ur Physik, Humboldt--Universit\"at zu Berlin,
Berlin, Germany\\ 
${}^b$ Russian Academy of Sciences, Scientific Council for
Cybernetics, Moscow, Russia\\

\end{center}

\begin{abstract}
Semileptonic and nonleptonic decays of the $B_c$ meson are studied
in the framework of the relativistic quark 
model. The heavy quark expansion in inverse powers of the heavy
($b,c$) quark mass is used to simplify
calculations while the light final  quarks in the $B$ and $D$ mesons are
treated relativistically. The decay form factors are explicitly 
expressed through the overlap integrals of the meson wave functions
in the whole accessible kinematical range. The
obtained  results are compared with the predictions of other
approaches.  
\end{abstract}

\section{Introduction}
                      
The investigation of weak decays of mesons composed of a heavy quark and
antiquark gives a very important insight in the heavy quark dynamics.
The properties of the $B_c$ meson are of special interest, since it
is the only heavy meson consisting of two heavy quarks with different
flavor. This difference of quark flavors forbids annihilation into
gluons. As a result, the excited $B_c$ meson states lying below the
$BD$ production threshold undergo pionic or radiative transitions to
the pseudoscalar ground state which is considerably more
stable than corresponding charmonium or bottomonium states and  decays
only weakly.  The
Collider Detector at Fermilab (CDF) Collaboration \cite{cdfcol}
reported the discovery of the $B_c$ ground state in $p\bar p$
collisions. More experimental data are expected to come in near future
from the Tevatron and Large Hadron Collider (LHC).

The characteristic feature of the $B_c$ meson  is that both
quarks forming it are heavy and thus their weak decays give comparable
contributions to the total decay rate. Therefore it is necessary to
consider both the $b$ quark decays $b\to c,u$ with the $\bar c$ quark
being a spectator and $\bar c$ quark decays $\bar c\to \bar s,\bar d$
with $b$ quark being a
spectator. The former transitions lead to semileptonic decays to
charmonium and $D$ mesons while the latter lead to decays to $B_s$ and
$B$ mesons. 
The estimates of the $B_c$ decay rates indicate that the $c$ quark decays
give the dominant contribution ($\sim 70\%$) while the $b$ quark
decays and weak annihilation contribute about 20\% and 10\%,
respectively (for a recent review see e.g. \cite{gklry} and references
therein). However, from the experimental point of view the $B_c$
decays to charmonium are easier to identify. Indeed, CDF observed
$B_c$ mesons \cite{cdfcol} analysing their semileptonic
decays  $B_c\to J/\psi l\nu$. 

The important difference between the $B_c$ semileptonic decays induced by
$b\to c,u$ and $c\to s,d$ transitions lies in the substantial
difference of their kinematical ranges. In the case of $B_c$ decays to
charmonium and $D^{(*)}$ mesons the kinematical range (the square of 
momentum transfer to the lepton pair varies from 0 to $q^2_{\rm max}\approx
10$~GeV$^2$ for decays to $J/\psi$ and $q^2_{\rm max}\approx
18$~GeV$^2$ for decays to $D$
mesons)  is considerably broader than for decays to $B_s^{(*)}$  and
$B^{(*)}$ mesons ($q^2_{\rm max}\approx 0.8$~GeV$^2$ for decays to $B_s$ and
$q^2_{\rm max}\approx 1$~GeV$^2$ for decays to $B$ mesons). As a
result in the 
$B_c$ meson rest frame the maximum recoil momentum of the final
charmonium and $D$ mesons is of the same order of magnitude as their
masses, while the maximum recoil momentum of the $B_s^{(*)}$ and
$B^{(*)}$ mesons is considerably smaller than the meson masses.   

In this talk we consider semileptonic $B_c$ decays  in the framework
of the relativistic quark model based on 
the quasipotential approach in quantum field theory. This model has
been successfully applied for the calculations of mass spectra,
radiative and weak decays of heavy quarkonia and heavy-light mesons
\cite{mass1,mass,hlm,gf,fg,mod}. In our recent
paper \cite{efgbc} we applied this model for the investigation of
properties of the $B_c$ meson and heavy quarkonia. The relativistic
wave functions obtained there are used for the calculation of the
transition matrix elements. The consistent theoretical description of
$B_c$ decays  requires a reliable determination of the
$q^2$ dependence of the decay amplitudes in the whole
kinematical range. In most previous calculations the
corresponding decay form factors were determined only at one
kinematical point either $q^2=0$ or $q^2=q^2_{\rm max}$ and then
extrapolated to the allowed kinematical range using some
phenomenological ansatz (mainly (di)pole or Gaussian). Our aim is to
explicitly  determine the $q^2$ dependence of form factors in the
whole kinematical range in order to avoid extrapolations thus reducing
uncertainties. 

 \section{Relativistic quark model}  
\label{rqm}

In the quasipotential approach a meson is described by the wave
function of the bound quark-antiquark state, which satisfies the
quasipotential equation \cite{3} of the Schr\"odinger type \cite{4}
\begin{equation}
\label{quas}
{\left(\frac{b^2(M)}{2\mu_{R}}-\frac{{\bf
p}^2}{2\mu_{R}}\right)\Psi_{M}({\bf p})} =\int\frac{d^3 q}{(2\pi)^3}
 V({\bf p,q};M)\Psi_{M}({\bf q}),
\end{equation}
where the relativistic reduced mass is
\begin{equation}
\mu_{R}=\frac{E_1E_2}{E_1+E_2}=\frac{M^4-(m^2_1-m^2_2)^2}{4M^3},
\end{equation}
and $E_1$, $E_2$ are the center of mass energies on mass shell given by
\begin{equation}
\label{ee}
E_1=\frac{M^2-m_2^2+m_1^2}{2M}, \quad E_2=\frac{M^2-m_1^2+m_2^2}{2M}.
\end{equation}
Here $M=E_1+E_2$ is the meson mass, $m_{1,2}$ are the quark masses,
and ${\bf p}$ is their relative momentum.  
In the center of mass system the relative momentum squared on mass shell 
reads
\begin{equation}
{b^2(M) }
=\frac{[M^2-(m_1+m_2)^2][M^2-(m_1-m_2)^2]}{4M^2}.
\end{equation}

The kernel 
$V({\bf p,q};M)$ in Eq.~(\ref{quas}) is the quasipotential operator of
the quark-antiquark interaction. It is constructed with the help of the
off-mass-shell scattering amplitude, projected onto the positive
energy states. 
Constructing the quasipotential of the quark-antiquark interaction, 
we have assumed that the effective
interaction is the sum of the usual one-gluon exchange term and the mixture
of long-range vector and scalar linear confining potentials, where
the vector confining potential
contains the Pauli interaction. The quasipotential is then defined by
\cite{mass1}
  \begin{equation}
\label{qpot}
V({\bf p,q};M)=\bar{u}_1(p)\bar{u}_2(-p){\mathcal V}({\bf p}, {\bf
q};M)u_1(q)u_2(-q),
\end{equation}
with
$${\mathcal V}({\bf p},{\bf q};M)=\frac{4}{3}\alpha_sD_{ \mu\nu}({\bf
k})\gamma_1^{\mu}\gamma_2^{\nu}
+V^V_{\rm conf}({\bf k})\Gamma_1^{\mu}
\Gamma_{2;\mu}+V^S_{\rm conf}({\bf k}),$$
where $\alpha_s$ is the QCD coupling constant, $D_{\mu\nu}$ is the
gluon propagator 
and ${\bf k=p-q}$; $\gamma_{\mu}$ and $u(p)$ are 
the Dirac matrices and spinors.
The effective long-range vector vertex is
given by
\begin{equation}
\label{kappa}
\Gamma_{\mu}({\bf k})=\gamma_{\mu}+
\frac{i\kappa}{2m}\sigma_{\mu\nu}k^{\nu},
\end{equation}
where $\kappa$ is the Pauli interaction constant characterizing the
long-range anomalous chromomagnetic moment of quarks. Vector and
scalar confining potentials in the nonrelativistic limit reduce to
\begin{equation}
\label{vlin}
V_V(r)=(1-\varepsilon)Ar+B,\qquad 
V_S(r) =\varepsilon Ar,
\end{equation}
reproducing 
\begin{equation}
\label{nr}
V_{\rm conf}(r)=V_S(r)+V_V(r)=Ar+B,
\end{equation}
where $\varepsilon$ is the mixing coefficient. 

The expression for the quasipotential of the heavy quarkonia,
expanded in $v^2/c^2$ without and with retardation corrections to the
confining potential, can be found in Refs.~\cite{mass1} and
\cite{efgbc,mass}, 
respectively. The 
structure of the spin-dependent interaction is in agreement with
the parameterization of Eichten and Feinberg \cite{ef}. The
quasipotential for the interaction of a heavy quark with a light antiquark
without employing the expansion in inverse powers of the light quark
mass is given in Ref.~\cite{hlm}.  
All the parameters of
our model like quark masses, parameters of the linear confining potential
$A$ and $B$, mixing coefficient $\varepsilon$ and anomalous
chromomagnetic quark moment $\kappa$ are fixed from the analysis of
heavy quarkonium masses \cite{mass1} and radiative
decays \cite{gf}. The quark masses
$m_b=4.88$ GeV, $m_c=1.55$ GeV, $m_s=0.50$ GeV, $m_{u,d}=0.33$ GeV and
the parameters of the linear potential $A=0.18$ GeV$^2$ and $B=-0.16$ GeV
have usual values of quark models.  The value of the mixing
coefficient of vector and scalar confining potentials $\varepsilon=-1$
has been determined from the consideration of the heavy quark expansion
for the semileptonic $B\to D$ decays
\cite{fg} and charmonium radiative decays \cite{gf}.
Finally, the universal Pauli interaction constant $\kappa=-1$ has been
fixed from the analysis of the fine splitting of heavy quarkonia ${
}^3P_J$- states \cite{mass1}. Note that the 
long-range  magnetic contribution to the potential in our model
is proportional to $(1+\kappa)$ and thus vanishes for the 
chosen value of $\kappa=-1$. It has been known for a long time 
that the correct reproduction of the
spin-dependent quark-antiquark interaction requires 
either assuming  the scalar confinement or equivalently  introducing the
Pauli interaction with $\kappa=-1$ \cite{schn,mass1,mass} in the vector
confinement.

\section{Matrix elements of the electroweak current} \label{mml}

In order to calculate the exclusive semileptonic decay rate of the
$B_c$ meson, it is necessary to determine the corresponding matrix
element of the  weak current between meson states.
In the quasipotential approach,  the matrix element of the weak current
$J^W_\mu=\bar q\gamma_\mu(1-\gamma_5)Q$, associated with $b\to q$
($Q=b$ and $q=c,u$) or $c\to q$ ($Q=c$ and $q=s,d$) transitions,
between a $B_c$ meson with mass $M_{B_c}$ and 
momentum $p_{B_c}$ and a final meson $F$ ($F=\psi, D, B_s, B$) with
mass $M_F$ and momentum $p_F$ takes the form \cite{f} 
\begin{equation}\label{mxet} 
\langle F(p_F) \vert J^W_\mu \vert B_c(p_{B_c})\rangle
=\int \frac{d^3p\, d^3q}{(2\pi )^6} \bar \Psi_{F\,{\bf p}_F}({\bf
p})\Gamma _\mu ({\bf p},{\bf q})\Psi_{B_c\,{\bf p}_{B_c}}({\bf q}),
\end{equation}
where $\Gamma _\mu ({\bf p},{\bf
q})$ is the two-particle vertex function and  
$\Psi_{M\,{\bf p}_M}$ are the
meson ($M=B_c,F)$ wave functions projected onto the positive energy
states of
quarks and boosted to the moving reference frame with momentum ${\bf p}_M$.

\begin{figure}
  \centering
  \includegraphics[height=3.7cm]{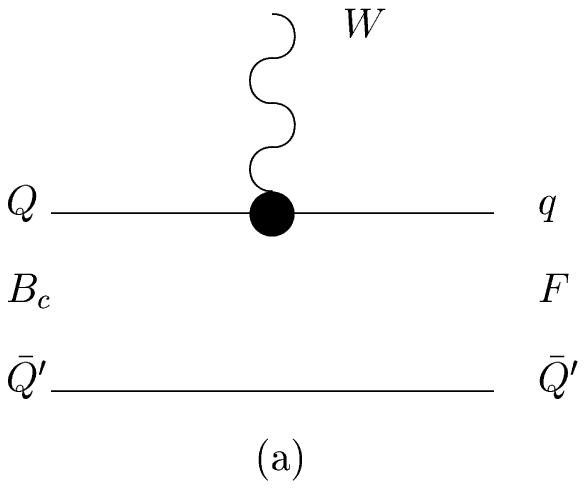}\qquad\qquad   
\includegraphics[height=3.7cm]{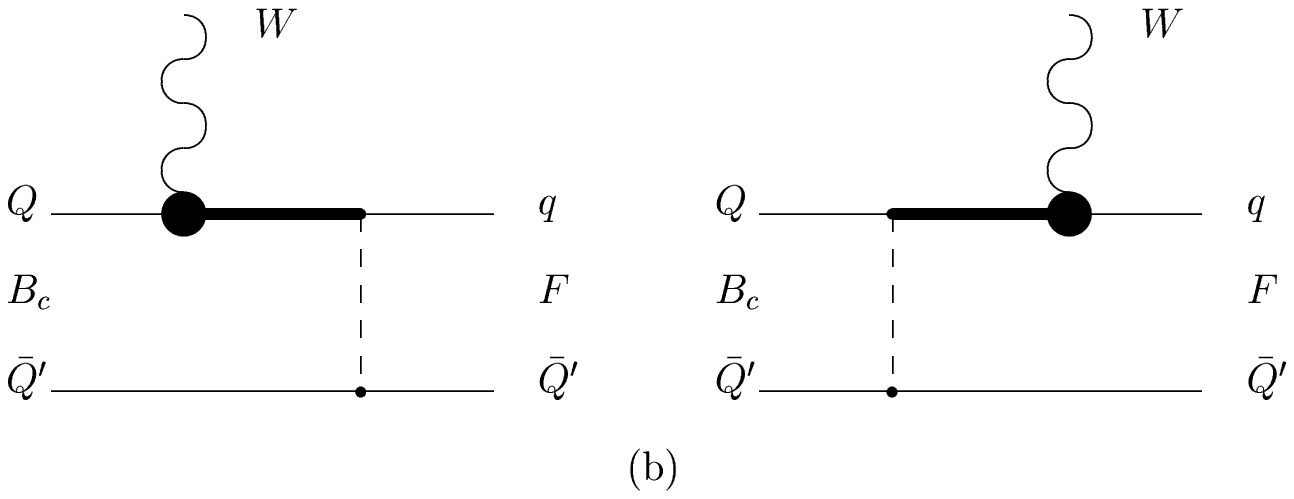}
\caption{(a) Lowest order vertex function $\Gamma^{(1)}$
contributing to the current matrix element (\ref{mxet}). \label{d1}
(b) Vertex function $\Gamma^{(2)}$
taking the quark interaction into account. Dashed lines correspond  
to the effective potential ${\cal V}$ in 
(\ref{qpot}). Bold lines denote the negative-energy part of the quark
propagator. }
\end{figure}

 The contributions to $\Gamma$ come from Figs.~\ref{d1}a,~\ref{d1}b. 
The contribution $\Gamma^{(2)}$ is the consequence
of the projection onto the positive-energy states. Note that the form of the
relativistic corrections resulting from the vertex function
$\Gamma^{(2)}$ is explicitly dependent on the Lorentz structure of the
quark-antiquark interaction. In the leading order of the $v^2/c^2$
expansion for heavy quarkonia and in the heavy quark limit $m_{b,c}\to
\infty$  for heavy-light mesons
only $\Gamma^{(1)}$ contributes, while $\Gamma^{(2)}$  
contributes already at the subleading order. 
The vertex functions look like
\begin{equation} \label{gamma1}
\Gamma_\mu^{(1)}({\bf
p},{\bf q})=\bar u_{q}(p_q)\gamma_\mu(1-\gamma^5)u_Q(q_Q)
(2\pi)^3\delta({\bf p}_{Q'}-{\bf
q}_{Q'}),\end{equation}
and
\begin{eqnarray}\label{gamma2} 
\Gamma_\mu^{(2)}({\bf
p},{\bf q})&=&\bar u_{q}(p_q)\bar u_{Q'}(p_{Q'})
\Bigl\{\gamma_{1\mu}(1-\gamma_1^5) 
\frac{\Lambda_Q^{(-)}(
k)}{\epsilon_Q(k)+\epsilon_Q(p_q)}\gamma_1^0
{\cal V}({\bf p}_{Q'}-{\bf
q}_{Q'})\nonumber \\ 
& &+{\cal V}({\bf p}_{Q'}-{\bf
q}_{Q'})\frac{\Lambda_{q}^{(-)}(k')}{ \epsilon_{q}(k')+
\epsilon_{q}(q_Q)}\gamma_1^0 \gamma_{1\mu}(1-\gamma_1^5)\Bigr\}u_Q(q_Q)
u_{Q'}(q_{Q'}),\end{eqnarray}
where the superscripts ``(1)" and ``(2)" correspond to Figs.~\ref{d1}a and
\ref{d1}b,  ${\bf k}={\bf p}_q-{\bf\Delta};\
{\bf k}'={\bf q}_Q+{\bf\Delta};\ {\bf\Delta}={\bf
p}_F-{\bf p}_{B_c}$;
$$\Lambda^{(-)}(p)=\frac{\epsilon(p)-\bigl( m\gamma
^0+\gamma^0({\bm{ \gamma}{\bf p}})\bigr)}{ 2\epsilon (p)}.$$
Here \cite{f} 
\begin{eqnarray*} 
p_{q,Q'}&=&\epsilon_{q,Q'}(p)\frac{p_F}{M_F}
\pm\sum_{i=1}^3 n^{(i)}(p_F)p^i,\\
q_{Q,Q'}&=&\epsilon_{Q,Q'}(q)\frac{p_{B_c}}{M_{B_c}} \pm \sum_{i=1}^3
n^{(i)} 
(p_{B_c})q^i,\end{eqnarray*}
and $n^{(i)}$ are three four-vectors given by
$$ n^{(i)\mu}(p)=\left\{ \frac{p^i}{M},\ \delta_{ij}+
\frac{p^ip^j}{M(E+M)}\right\}, \quad E=\sqrt{{\bf p}^2+M^2}.$$

The general structure of the current matrix element (\ref{mxet}) is
rather complicated, because it is necessary to integrate both with
respect to $d^3p$ and $d^3q$. The $\delta$-function in the expression
(\ref{gamma1}) for the vertex function $\Gamma^{(1)}$ permits to perform
one of these integrations. As a result the contribution of
$\Gamma^{(1)}$ to the current matrix element has the usual structure of
an overlap integral of meson wave functions and
can be calculated exactly (without employing any expansion) in the
whole kinematical range, if the wave functions of the
initial and final meson are known. The situation with the contribution
$\Gamma^{(2)}$ is different. Here, instead of a $\delta$-function, we have
a complicated structure, containing the potential of the $q\bar
q$-interaction in meson. Thus in the general case we cannot get rid of one
of the integrations in the contribution of $\Gamma^{(2)}$ to the
matrix element (\ref{mxet}). Therefore, it is necessary to use some 
additional considerations in order to simplify calculations. The main
idea is to expand the vertex 
function $\Gamma^{(2)}$, given by (\ref{gamma2}), in such  a way that it
will be possible to use the quasipotential equation (\ref{quas}) in order
to perform one of the integrations in the current matrix element
(\ref{mxet}).  

{ \bf  $\bm{B_c\to \psi,\eta_c}$ transitions.} The
natural expansion parameters for $B_c$ decays to charmonium are the
active heavy $b$ and $c$ quark masses as well as the spectator $c$
quark mass. We carry such an expansion up to the second order in the ratios
of the relative quark momentum ${\bf p}$ and binding energy to the
heavy quark masses $m_{b,c}$. It is important to take into account the
fact that in the case of weak $B_c$ decays caused by $b\to c,u$ quark
transition the kinematically allowed range is large
($|{\bf\Delta}_{\rm max}|=(M_{B_c}^2-M_F^2)/(2M_{B_c})$ $\sim 2.4$~GeV for
decays to charmonium and $\sim 2.8$~GeV for decays to $D$ mesons). This
means that the recoil momentum ${\bf \Delta}$ of a final meson is large in
comparison to the relative momentum ${\bf p}$ of quarks inside a meson 
($\sim 0.5$~GeV), being of the same order as the heavy
quark mass  almost in the whole kinematical range. Thus we do not use
expansions in powers of $|{\bf\Delta}|/m_{b,c}$ or
$|{\bf\Delta}|/M_F$, but approximate in the expression (\ref{gamma2}) for
$\Gamma^{(2)}$ the heavy
quark energies $\epsilon_{b,c}(p+\Delta)\equiv\sqrt{m_{b,c}^2+({\bf
p}+{\bf\Delta})^2}$ by $\epsilon_{b,c}(\Delta)\equiv
\sqrt{m_{b,c}^2+{\bf\Delta}^2}$, which become independent of the quark
relative momentum 
${\bf p}$. Making these replacements and expansions we see that it is
possible to 
integrate the current matrix element (\ref{mxet}) either with
respect to $d^3p$ or $d^3q$ using the quasipotential equation
(\ref{quas}). Performing integrations and
taking the sum of the contributions of $\Gamma^{(1)}$ and
$\Gamma^{(2)}$ we get the expression for the current matrix element,
which contains ordinary overlap integrals of meson wave functions
and is valid in the whole kinematical range.
Thus this matrix element can be easily calculated using numerical wave
functions found in our meson mass spectrum analysis \cite{efgbc,mass}.

{ \bf  $\bm{B_c\to D^{(*)}}$ transitions.} In this case the heavy
$b$ quark undergoes the weak transition to the light $u$ quark. The
constituent $u$ quark mass is of the same order of magnitude as the relative
momentum and binding energy, thus we cannot apply the expansion in
inverse powers of its mass. Nevertheless, taking into account the fact
that the recoil momentum of the final meson in this decay is large
almost in the whole kinematical range (as it was discussed above), we
can neglect the relative momentum ${\bf p}$ of quarks inside a meson
with respect to the large recoil momentum ${\bf \Delta}$. Thus in the
region of large recoil ($|{\bf \Delta}|\gg |{\bf p}| $) we can
use the same expressions of the $\Gamma^{(2)}$ contribution to the
current matrix element both for the $B_c\to D^{(*)}$ and
$B_c\to\psi,\eta_c$  transitions. Moreover, the smallness of
the $\Gamma^{(2)}$ contribution, 
which is proportional to the small binding energy, and its weak
dependence on momentum transfer allows one to extrapolate these
formulae to the whole kinematical range. As numerical estimates show,
such extrapolation introduces only small uncertainties.

{ \bf  $\bm{B_c\to B_s(B)^{(*)}}$ transitions.} The
heavy $c$ quark undergoes the weak transition to the light $s$ or $d$
quark. The constituent  $s,d$ quark masses are of the same order of
magnitude as the relative momentum and binding energy, thus we cannot
apply the expansion in inverse powers of their masses. The heavy quark
expansion in $1/m_{c,b}$ significantly simplifies the structure of
the $\Gamma^{(2)}$ contribution to the decay matrix element, but the
momentum ${\bf p}$ dependence of the light quark energies
$\epsilon_q(p)$ still prevents to perform one of the
integrations. It is important to note that the kinematically allowed
range for $B_c$ decays to $B_s$ and $B$ meson is not large 
($|{\bf\Delta}_{\rm max}|=(M_{B_c}^2-M_F^2)/(2M_{B_c})$ $\sim 0.8$~GeV for
decays to $B_s$ and $\sim 0.9$~GeV for decays to $B$ mesons). This
means that the recoil momentum ${\bf \Delta}$ of a final meson is
of the same order as the relative momentum ${\bf p}$ of
quarks inside a heavy-light meson ($\sim 0.5$~GeV) in the whole
kinematical range. Taking also into account that the
final $B_s$ and $B$ mesons are weakly bound \cite{hlm}, 
we can replace the light quark
energies by the center of mass energies on mass shell
$\epsilon_q(p)\to E_q=(M_F^2-m_b^2+m_q^2)/(2M_F)$. We used such a
substitution in our analysis of heavy-light meson mass spectra
\cite{hlm} which allowed us to treat the light quark relativistically
without an unjustified expansion in inverse powers of its mass. Making
these replacements and expansions we see that it is possible to 
integrate the current matrix element (\ref{mxet}) either with
respect to $d^3p$ or $d^3q$ using the quasipotential equation
(\ref{quas}). Performing integrations and
taking the sum of the contributions $\Gamma^{(1)}$ and
$\Gamma^{(2)}$ we get the expression for the current matrix element,
which contains ordinary overlap integrals of meson wave functions
and is valid in the whole kinematical range.

\section{$\bm{B_{\lowercase{c}}}$ decay form factors and weak decay
  rates} \label{dff}

 The matrix elements of the weak current $J^W$ for $B_c$ decays
 to pseudoscalar  mesons
$P$ can be parametrized by two invariant form factors:
\begin{equation}
  \label{eq:pff1}
  \langle P(p_F)|\bar q \gamma^\mu b|B_c(p_{B_c})\rangle
  =f_+(q^2)\left[p_{B_c}^\mu+ p_F^\mu-
\frac{M_{B_c}^2-M_P^2}{q^2}\ q^\mu\right]+
  f_0(q^2)\frac{M_{B_c}^2-M_P^2}{q^2}\ q^\mu,
\end{equation}
where $q=p_{B_c}-p_F$; $M_{B_c}$ is the $B_c$ meson mass 
and $M_P$ is the pseudoscalar meson mass. 

The corresponding matrix elements for $B_c$  decays to vector mesons
$V$ are parametrized by four form factors
\begin{eqnarray}
  \label{eq:vff1}
  \langle V(p_F)|\bar q \gamma^\mu b|B(p_{B_c})\rangle&=
  &\frac{2iV(q^2)}{M_{B_c}+M_V} \epsilon^{\mu\nu\rho\sigma}\epsilon^*_\nu
  p_{B_c\rho} p_{F\sigma},\\ \cr
\label{eq:vff2}
\langle V(p_F)|\bar q \gamma^\mu\gamma_5 b|B(p_{B_c})\rangle&=&2M_V
A_0(q^2)\frac{\epsilon^*\cdot q}{q^2}\ q^\mu
 +(M_{B_c}+M_V)A_1(q^2)\left(\epsilon^{*\mu}-\frac{\epsilon^*\cdot
    q}{q^2}\ q^\mu\right)\cr\cr
&&-A_2(q^2)\frac{\epsilon^*\cdot q}{M_{B_c}+M_V}\left[p_{B_c}^\mu+
  p_F^\mu-\frac{M_{B_c}^2-M_V^2}{q^2}\ q^\mu\right], 
\end{eqnarray}
where 
$M_V$ and $\epsilon_\mu$ are the mass and polarization vector of
the final vector meson. The following relations hold for the form
factors at the maximum recoil point of the final meson ($q^2=0$)
\[f_+(0)=f_0(0),\]
\[A_0(0)=\frac{M_{B_c}+M_V}{2M_V}A_1(0)
-\frac{M_{B_c}-M_V}{2M_V}A_2(0).\]
In the limit of vanishing lepton mass, the form factors $f_0$ and $A_0$ do
not contribute to the semileptonic decay rates. However, they
contribute to nonleptonic decay rates in the factorization
approximation.

\begin{table}
\caption{Form factors of weak $B_c$ decays. }
\label{ff}
\begin{center}
\begin{tabular}{ccccccc}
\hline
Transition   & $f_+(q^2)$ & $f_0(q^2)$ & $V(q^2)$ & $A_1(q^2)$ &
$A_2(q^2)$ & 
   $A_0(q^2)$\\
\hline
$B_c\to\eta_c,J/\psi$ & \\
$q^2=q^2_{\rm max}$ & 1.07 & 0.92 & 1.34 & 0.88 & 1.33 & 1.06\\  
$q^2=0$ & 0.47 & 0.47 & 0.49 & 0.50 & 0.73 & 0.40\\
$B_c\to\eta_c',\psi'$ & \\
$q^2=q^2_{\rm max}$ & 0.08 & 0.05 & $-0.16$ & 0.03 & 0.10 & 0.08\\
$q^2=0$ & 0.27 & 0.27 & 0.24 & 0.18 & 0.14 & 0.23\\
$B_c\to D,D^*$ &\\
$q^2=q^2_{\rm max}$ & 1.20 & 0.64 & 2.60 & 0.62 & 1.78 & 0.97   \\
$q^2=0$ & 0.14 & 0.14 & 0.18 & 0.17 & 0.19 & 0.14 \\
$B_c\to B_s(B_s^*)$ & \\
$q^2=q^2_{\rm max}$ & 0.99 & 0.86 & 6.25 & 0.76 & 2.62 & 0.91\\  
$q^2=0$ & 0.50 & 0.50 & 3.44 & 0.49 & 2.19 & 0.35\\
$B_c\to B(B^*)$ &\\
$q^2=q^2_{\rm max}$ & 0.96 & 0.80 & 8.91 & 0.72 & 2.83 & 1.06   \\
$q^2=0$ & 0.39 & 0.39 & 3.94 & 0.42 & 2.89 & 0.20 \\
\hline
\end{tabular}
\end{center}
\end{table}

Now we apply the method for the calculation of the decay matrix
elements described in the previous section. The explicit expressions
for the form factors can be found in Refs.~\cite{wbc1,wbc2}.
Their calculated values are given in Table~\ref{ff}.

The differential semileptonic decay rates can be expressed in terms of
the form factors as follows.

(a) $B_c\to Pe\nu$ decays  
\begin{equation}
  \label{eq:dgp}
  \frac{{\rm d}\Gamma}{{\rm d}q^2}(B_c\to Pe\nu)=\frac{G_F^2 
  \Delta^3 |V_{qQ}|^2}{24\pi^3} |f_+(q^2)|^2.
\end{equation}

(b) $B_c\to Ve\nu$ decays 
\begin{equation}
  \label{eq:dgv}
\frac{{\rm d}\Gamma}{{\rm d}q^2}(B_c\to Ve\nu)=\frac{G_F^2
\Delta|V_{qQ}|^2}{96\pi^3}\frac{q^2}{M_{B_c}^2}
\left(|H_+(q^2)|^2+|H_-(q^2)|^2   
+|H_0(q^2)|^2\right),
\end{equation}
where $G_F$ is the Fermi constant, $V_{qQ}$ is the
CKM matrix element,
\[\Delta\equiv|{\bf\Delta}|=\sqrt{\frac{(M_{B_c}^2+M_{P,V}^2-q^2)^2}
{4M_{B_c}^2}-M_{P,V}^2}.
\]
The helicity amplitudes are given by
\begin{equation}
  \label{eq:helamp}
  H_\pm(q^2)=\frac{2M_{B_c}\Delta}{M_{B_c}+M_V}\left[V(q^2)\mp
\frac{(M_{B_c}+M_V)^2}{2M_{B_c}\Delta}A_1(q^2)\right],
\end{equation}
\begin{equation}
  \label{eq:h0a}
  H_0(q^2)=\frac1{2M_V\sqrt{q^2}}\left[(M_{B_c}+M_V)
(M_{B_c}^2-M_V^2-q^2)A_1(q^2)-\frac{4M_{B_c}^2\Delta^2}{M_{B_c}
+M_V}A_2(q^2)\right].
\end{equation}

\begin{table}
\caption{Semileptonic decay rates $\Gamma$ (in $10^{-15}$ GeV) of
  $B_c$ decays. } 
\label{ssdr}
\begin{tabular}{@{}c@{ }cccccccccc@{}}
\hline
Decay& our& \cite{iks}& \cite{klo} & \cite{emv} & \cite{cc} &
\cite{cdf} & \cite{aknt} & \cite{nw} &\cite{lyl}&\cite{lc}\\
\hline
$B_c\to\eta_c e\nu$ & 5.9 & 14 & 11 & 11.1 & 14.2 & 2.1(6.9) & 8.6 &
6.8 & 4.3& 8.31\\
$B_c\to\eta_c' e\nu$ & 0.46 &  & 0.60 &  & 0.73 & 0.3 & & &  & 0.605\\
$\!\!B_c\to J/\psi e\nu$ & 17.7 & 33 & 28 & 30.2 & 34.4 & 21.6(48.3) &
17.5 & 19.4 & 16.8 & 20.3\\
$B_c\to\psi' e\nu$ & 0.44 &  & 1.94 &  & 1.45 & 1.7 & & & & 0.186\\
$B_c\to D e\nu$ & 0.019 & 0.26 & 0.059 & 0.049 & 0.094 & 0.005(0.03) &
& & 0.001 & $\!\!$0.0853\\
$B_c\to D^* e\nu$ & 0.11 & 0.49 & 0.27 & 0.192 & 0.269 & 0.12(0.5) & &
& 0.06 & 0.204\\
$B_c\to B_s e\nu$ & 12 & 29 & 59 & 14.3 & 26.6 & 11.1(12.9) & 15 &
12.3 & 11.75 & 26.8\\
$B_c\to B_s^* e\nu$ & 25 & 37 & 65 & 50.4 & 44.0 & 33.5(37.0) &
34 & 19.0 & 32.56 & 34.6\\
$B_c\to B e\nu$ & 0.6 & 2.1 & 4.9 & 1.14 & 2.30 & 0.9(1.0) &
& & 0.59 & 1.90\\
$B_c\to B^* e\nu$ & 1.7 & 2.3 & 8.5 & 3.53 & 3.32 & 2.8(3.2) & &
& 2.44 & 2.34\\
\hline
\end{tabular}
\end{table}

\begin{table}
\caption{Nonleptonic $B_c$ decay rates $\Gamma$ (in $10^{-15}$ GeV).}
\label{nldr}
\begin{tabular}{@{}cccccccc@{}}
\hline
Decay& our&  \cite{klo} & \cite{emv} & \cite{cc} & \cite{aknt} &
\cite{cdf}& \cite{lc} \\
\hline
$B_c^+\to\eta_c\pi^+$ & 0.93$a_1^2$ & 1.8$a_1^2$ & 1.59$a_1^2$ &
2.07$a_1^2$ & 1.47$a_1^2$ & 0.28$a_1^2$ & 1.49$a_1^2$ \\
$B_c^+\to\eta_c\rho^+$ & 2.3$a_1^2$ & 4.5$a_1^2$ & 3.74$a_1^2$ &
5.48$a_1^2$ & 3.35$a_1^2$ & 0.75$a_1^2$ & 3.93$a_1^2$\\
$B_c^+\to J/\psi\pi^+$ & 0.67$a_1^2$ & 1.43$a_1^2$ & 1.22$a_1^2$ &
1.97$a_1^2$ & 0.82$a_1^2$ & 1.48$a_1^2$& 1.01$a_1^2$\\
$B_c^+\to J/\psi\rho^+$ & 1.8$a_1^2$ & 4.37$a_1^2$ & 3.48$a_1^2$ &
5.95$a_1^2$ & 2.32$a_1^2$ & 4.14$a_1^2$ & 3.25$a_1^2$\\
$B_c^+\to\eta_c K^+$ & 0.073$a_1^2$ & 0.15$a_1^2$ & 0.119$a_1^2$ &
0.161$a_1^2$ & 0.15$a_1^2$ & 0.023$a_1^2$ & 0.115$a_1^2$\\
$B_c^+\to\eta_c K^{*+}$ & 0.12$a_1^2$ & 0.22$a_1^2$ & 0.200$a_1^2$ &
0.286$a_1^2$ & 0.24$a_1^2$ & 0.041$a_1^2$ & 0.198$a_1^2$\\
$B_c^+\to J/\psi K^+$ & 0.052$a_1^2$ & 0.12$a_1^2$ & 0.090 $a_1^2$ &
0.152$a_1^2$ & 0.079$a_1^2$ & 0.076$a_1^2$& 0.0764$a_1^2$\\
$B_c^+\to J/\psi K^{*+}$ & 0.11$a_1^2$ & 0.25$a_1^2$ & 0.197$a_1^2$ &
0.324$a_1^2$ & 0.18$a_1^2$ &0.23$a_1^2$ & 0.174$a_1^2$\\
$B_c^+\to\eta_c'\pi^+$ & 0.19$a_1^2$ & & & 0.268$a_1^2$ & &
0.074$a_1^2$ & 0.248$a_1^2$\\ 
$B_c^+\to\eta_c'\rho^+$ & 0.40$a_1^2$ & & & 0.622$a_1^2$ & &
0.16$a_1^2$& 0.587$a_1^2$\\
$B_c^+\to \psi'\pi^+$ & 0.12$a_1^2$ & & & 0.252$a_1^2$& &
0.22$a_1^2$& 0.0708$a_1^2$\\
$B_c^+\to \psi'\rho^+$ & 0.20$a_1^2$ & & & 0.710 $a_1^2$& &
0.54$a_1^2$& 0.183$a_1^2$\\
$B_c^+\to\eta_c'K^+$ & 0.014$a_1^2$ & & & 0.020$a_1^2$ & &
0.0055$a_1^2$& 0.0184$a_1^2$\\ 
$B_c^+\to\eta_c'K^{*+}$ & 0.021$a_1^2$ & & & 0.031$a_1^2$ & &
0.008$a_1^2$& 0.0283$a_1^2$\\
$B_c^+\to \psi'K^+$ & 0.009$a_1^2$ & & & 0.018$a_1^2$& &
0.01$a_1^2$& 0.00499$a_1^2$\\
$B_c^+\to \psi'K^{*+}$ & 0.011$a_1^2$ & & & 0.038 $a_1^2$& &
0.03$a_1^2$& 0.00909$a_1^2$\\
$B_c^+\to B_s\pi^+$ & 25$a_1^2$ & 167$a_1^2$ & 15.8$a_1^2$ &
58.4$a_1^2$ & 34.8$a_1^2$ & 30.6$a_1^2$ & 65.1$a_1^2$ \\
$B_c^+\to B_s\rho^+$ & 14$a_1^2$ & 72.5$a_1^2$ & 39.2$a_1^2$ &
44.8$a_1^2$ & 23.6$a_1^2$ & 13.6$a_1^2$& 42.7$a_1^2$\\
$B_c^+\to B_s^*\pi^+$ & 16$a_1^2$ & 66.3$a_1^2$ & 12.5$a_1^2$ &
51.6$a_1^2$ & 19.8$a_1^2$ & 35.6$a_1^2$& 25.3$a_1^2$\\
$B_c^+\to B_s^*\rho^+$ & 110$a_1^2$ & 204$a_1^2$ & 171$a_1^2$ &
150$a_1^2$ & 123$a_1^2$ & 110.1$a_1^2$ & 139.6$a_1^2$\\
$B_c^+\to B_s K^+$ & 2.1$a_1^2$ & 10.7$a_1^2$ & 1.70$a_1^2$ &
4.20$a_1^2$ &  & 2.15$a_1^2$ & 4.69$a_1^2$\\
$B_c^+\to B_s K^{*+}$ & 0.03$a_1^2$ &  & 1.06$a_1^2$ &
 &  & 0.043$a_1^2$ & 0.296$a_1^2$\\
$B_c^+\to B_s^* K^+$ & 1.1$a_1^2$ & 3.8$a_1^2$ & 1.34$a_1^2$ &
2.96$a_1^2$ &  & 1.6$a_1^2$ & 1.34$a_1^2$\\
$B_c^+\to B^0\pi^+$ & 1.0$a_1^2$ & 10.6$a_1^2$ & 1.03$a_1^2$ &
3.30$a_1^2$ & 1.50$a_1^2$ & 1.97$a_1^2$& 3.64$a_1^2$\\
$B_c^+\to B^0\rho^+$ & 1.3$a_1^2$ & 9.7$a_1^2$ & 2.81$a_1^2$ &
5.97$a_1^2$ & 1.93$a_1^2$ & 1.54$a_1^2$ & 4.03$a_1^2$\\
$B_c^+\to B^{*0}\pi^+$ & 0.26$a_1^2$ & 9.5$a_1^2$ & 0.77$a_1^2$ &
2.90$a_1^2$ & 0.78$a_1^2$ & 2.4$a_1^2$& 1.22$a_1^2$\\
$B_c^+\to B^{*0}\rho^+$ & 6.8$a_1^2$ & 26.1$a_1^2$ & 9.01$a_1^2$ &
11.9$a_1^2$ & 6.78$a_1^2$ & 8.6$a_1^2$ & 8.16$a_1^2$\\
$B_c^+\to B^0 K^+$ & 0.09$a_1^2$ & 0.70$a_1^2$ & 0.105$a_1^2$ &
0.255$a_1^2$ &  & 0.14$a_1^2$ & 0.272$a_1^2$\\
$B_c^+\to B^0 K^{*+}$ & 0.04$a_1^2$ & 0.15$a_1^2$ & 0.125$a_1^2$ &
0.180$a_1^2$ &  & 0.032$a_1^2$ & 0.0965$a_1^2$\\
$B_c^+\to B^{*0} K^+$ & 0.04$a_1^2$ & 0.56$a_1^2$ & 0.064$a_1^2$ &
0.195$a_1^2$ &  & 0.12$a_1^2$ & 0.0742$a_1^2$\\
$B_c^+\to B^{*0} K^{*+}$ & 0.33$a_1^2$ & 0.59$a_1^2$ & 0.665$a_1^2$ &
0.374$a_1^2$ &  & 0.34$a_1^2$ & 0.378$a_1^2$\\
$B_c^+\to B^+\bar K^0$ & 34$a_2^2$ & 286$a_2^2$ & 39.1$a_2^2$ &
96.5$a_2^2$ & 24.0$a_2^2$  & & 103.4$a_2^2$\\
$B_c^+\to B^+\bar K^{*0}$ & 13$a_2^2$ & 64$a_2^2$ & 46.8$a_2^2$ &
68.2$a_2^2$ & 13.8$a_2^2$  &  & 36.6$a_2^2$\\
$B_c^+\to B^{*+}\bar K^0$ & 15$a_2^2$ & 231$a_2^2$ & 24.0$a_2^2$ &
73.3$a_2^2$ & 8.9$a_2^2$  &  & 28.9$a_2^2$\\
$B_c^+\to B^{*+}\bar K^{*0}$ & 120$a_2^2$ & 242$a_2^2$ & 247$a_2^2$ &
141$a_2^2$ & 82.3$a_2^2$  & & 143.6$a_2^2$\\
$B_c^+\to B^+\pi^0$ & 0.5$a_2^2$ & 5.3$a_2^2$ & 0.51$a_2^2$ &
1.65$a_2^2$ & 1.03$a_2^2$ & & \\
$B_c^+\to B^+\rho^0$ & 0.7$a_2^2$ & 4.4$a_2^2$ & 1.40$a_2^2$ &
2.98$a_2^2$ & 1.28$a_2^2$ & & \\
$B_c^+\to B^{*+}\pi^0$ & 0.13$a_2^2$ & 4.8$a_2^2$ & 0.38$a_2^2$ &
1.45$a_2^2$ & 0.53$a_2^2$ & & \\
$B_c^+\to B^{*+}\rho^0$ & 3.4$a_2^2$ & 13.1$a_2^2$ & 4.50$a_2^2$ &
5.96$a_2^2$ & 4.56$a_2^2$ & &\\
\hline
\end{tabular}
 \centerline{$a_{1,2}$ are the Wilson coefficients in the operator
 product expansion.} 
\end{table}

The results of our calculations of the semileptonic and nonleptonic
(in the factorization approximation) decay rates of $B_c$ meson  are
given in Tables~\ref{ssdr}, \ref{nldr} in comparison with predictions
of other approaches based on 
quark models \cite{iks,emv,cc,aknt,nw,lc}, QCD sum
rules \cite{klo} and on the application of heavy quark
symmetry relations \cite{cdf,lyl} to the quark model. Our predictions
for the CKM favored semileptonic $B_c$ decays to charmonium ground states
are almost 2 times smaller than those of QCD sum rules \cite{klo} and
quark models \cite{iks,emv,cc}, but agree with quark model results
\cite{aknt,nw,lyl,lc}. Note that the ratios of the $B_c\to J/\psi
e\nu$ to $B_c\to \eta_c e\nu$ decay rates have close values in all
approaches except 
\cite{cdf}. In the case of semileptonic decays to radially excited
charmonium states our prediction for the decay to the pseudoscalar $\eta_c'$
state is consistent with others, while the one for the decay to
$\psi'$ is considerably smaller (with the exception of Ref.~\cite{lc}). For
the CKM suppressed 
semileptonic decays of $B_c$ to $D$ mesons our results are in agreement
with those of Ref.~\cite{cdf}. Our predictions
for the CKM favored semileptonic $B_c$ decays to $B_s^{(*)}$
are  smaller than those of QCD sum rules \cite{klo} and
quark models \cite{iks,emv,cc,lc}, but agree with quark model results
\cite{cdf,aknt,nw,lyl}. For the CKM suppressed
semileptonic decays of $B_c$ to $B^{(*)}$ mesons our results are in
agreement with the ones based on the application of heavy quark symmetry
relations \cite{cdf,lyl} to the quark model.

As one sees from Tables~\ref{ssdr},~\ref{nldr} the theoretical
predictions for $B_c$ weak decay rates differ substantially. Thus
experimental measurements of corresponding decay rates can
discriminate between various approaches.     

\section{Conclusions}

\begin{table}
\caption{Branching fractions (in \%) of exclusive $B_c$ decays calculated
  for the fixed values of the $B_c$ lifetime $\tau_{B_c}=0.46$~ps,
  $a_1=1.14$ for $b$-quark decays and $a_1=1.20$, $a_2=-0.317$ for
  $c$-quark decays.}\label{Br} 
\begin{center}
\begin{tabular}{cccccc}
\hline
Decay& Br & Decay & Br & Decay & Br\\
\hline
$b$-quark decays\\
$B_c\to \eta_c e\nu$& 0.42 &$B_c^+\to\eta_c\pi^+ $& 0.085  &
$B_c^+\to \eta' \pi^{+}$& 0.017 \\ 
$B_c\to \eta'_c e\nu$& 0.032 &$B_c^+\to\eta_c\rho^+ $& 0.21 & $B_c^+\to
\eta_c' \rho^{+}$& 0.036 \\
$B_c\to J/\psi e\nu$& 1.23 &$B_c^+\to J/\psi\pi^+ $& 0.061  &$B_c^+\to
\psi' \pi^{+}$& 0.011\\
 $B_c\to \psi' e\nu$& 0.031 &$B_c^+\to J/\psi \rho^+$& 0.16 &
 $B_c^+\to \psi' \rho^{+}$& 0.018 \\ 
$B_c\to D e\nu$& 0.013 & $B_c^+\to \eta_c K^+$& 0.007 &$B_c^+\to
\eta_c' K^+$& 0.001\\
$B_c\to D^* e\nu$& 0.037 &$B_c^+\to \eta_c K^{*+}$& 0.011&$B_c^+\to
\eta_c' K^{*+}$& 0.002\\
& & $B_c^+\to J/\psi K^{+}$& 0.005 & $B_c^+\to
\psi' K^{+}$& 0.001\\
& &$B_c^+\to J/\psi K^{*+}$& 0.010  &$B_c^+\to
\psi' K^{*+}$& 0.001\\
\hline
$c$-quark decays\\
$B_c\to B_s e\nu$& 0.84 &$B_c^+\to B_s K^{*+} $& 0.003  &
$B_c^+\to B^{*0} K^{*+}$& 0.033 \\ 
$B_c\to B^*_s e\nu$& 1.75 &$B_c^+\to B_s^* K^+ $& 0.11 & $B_c^+\to
B^+ \bar K^{0}$& 0.24 \\
$B_c\to B e\nu$& 0.042 &$B_c^+\to B^0\pi^+ $& 0.10  &$B_c^+\to
B^+ \bar K^{*0}$& 0.09\\
 $B_c\to B^* e\nu$& 0.12 &$B_c^+\to B^0 \rho^+$& 0.13 &
 $B_c^+\to B^{*+} \bar K^{0}$& 0.11 \\ 
$B_c^+\to B_s\pi^+$& 2.52 & $B_c^+\to B^{*0} \pi^+$& 0.026 &$B_c^+\to
B^{*+} \bar K^{*0}$& 0.84\\
$B_c^+\to B_s\rho^+$& 1.41 &$B_c^+\to B^{*0} \rho^{+}$& 0.68&$B_c^+\to
B^+ \pi^{0}$& 0.004\\
$B_c^+\to B_s^*\pi^+$&1.61 & $B_c^+\to B^0 K^{+}$& 0.009 & $B_c^+\to
B^+ \rho^{0}$& 0.005\\
$B_c^+\to B_s^*\rho^+$&11.1 &$B_c^+\to B^0 K^{*+}$& 0.004  &$B_c^+\to
B^{*+} \pi^{0}$& 0.001\\
$B_c^+\to B_s K^+$&0.21 &$B_c^+\to B^{*0} K^{+}$& 0.004  &$B_c^+\to
B^{*+} \rho^{0}$& 0.024\\
\hline
\end{tabular}
\end{center}
\end{table}
In this talk we considered weak semileptonic and nonleptonic $B_c$
decays in the framework of the relativistic
quark model based on the quasipotential approach in quantum field
theory. The weak decay form factors were calculated explicitly in the
whole kinematical range using the heavy quark expansion for the 
initial active heavy quark $Q=b,c$ and spectator antiquark $\bar Q'=\bar
c,\bar b$. The final light
quark  was treated completely relativistically without
applying the unjustified expansion in inverse powers of its mass. 
The leading order contribution of the heavy quark expansion was treated
exactly, while in calculating the subleading order contribution
additional simplifying replacements were used. It was shown
that such substitutions introduce only minor errors which are of
the same order as the higher order terms in the heavy quark expansion. Thus
the decay form factors were evaluated up to the subleading order of the
heavy quark expansion. The overall subleading contributions are small
and weakly depend on the momentum transfer $q^2$.

We calculated semileptonic and nonleptonic (in the factorization
approximation) $B_c$ decay rates. 
Our predictions for the branching fractions are summarized in
Table~\ref{Br}, where we use the central experimental value of the
$B_c$ meson lifetime \cite{pdg}. From this table we see that weak
$B_c$ semileptonic decays to the ground and first radially excited states
of charmonium and to $D$ mesons, associated 
with $\bar b\to \bar c, \bar u$ quark transition, yield $\sim1.7$\%
and corresponding energetic nonleptonic decays (to charmonium and
$K^{(*)}$ or $\pi,\rho$ mesons) contribute $\sim0.6$\%.
The semileptonic 
decays to $B_s$ and $B$ mesons, associated 
with $c\to s,  d$ quark transition, give in total $\sim2.0$\% of the $B_c$
decay rate, while the energetic nonleptonic decays provide
the dominant contribution $\sim19.3$\%. All these decays (to $B_s$, $B$,
charmonium and $D$ mesons) 
add up to $\sim 23.6$\% of the $B_c$ total decay rate.

The authors express their gratitude to V. Kiselev, J. K\"orner,
M. M\"uller-Preussker, V. Papadimitriou, V. Savrin and A. Vairo for
support and discussions. 
Two of us (R.N.F and V.O.G.) were supported in part by the 
{\it Deutsche Forschungsgemeinschaft} under contract Eb 139/2-2.



\begin{thebibliography}{99}
\bibitem{cdfcol} CDF Collaboration, F. Abe {\it et al.}, Phys. Rev. D
  {\bf 58}, 112004 (1998).
 \vspace{-2.5mm}
\bibitem{gklry} I. P. Gouz, V. V. Kiselev, A. K. Likhoded,
  V. I. Romanovsky, and O. P. Yushchenko, hep-ph/0211432.
\vspace{-2.5mm}
\bibitem{mass1} V. O. Galkin, A. Yu. Mishurov and R. N. Faustov, Yad. Fiz.
{\bf 55}, 2175 (1992) [Sov. J. Nucl. Phys. {\bf 55}, 1207 (1992)].
\vspace{-2.5mm}
\bibitem{mass}  D. Ebert, R. N. Faustov and V. O. Galkin, Phys. Rev. D
  {\bf 62}, 034014 (2000).
\vspace{-2.5mm}
\bibitem{hlm} D. Ebert, V. O. Galkin and R. N. Faustov, Phys. Rev. D
  {\bf 57}, 5663 (1998); {\bf 59}, 019902(E) (1999).
\vspace{-2.5mm}
\bibitem{gf} V. O. Galkin and R. N. Faustov, Yad. Fiz. {\bf 44}, 1575
(1986) [Sov. J. Nucl. Phys. {\bf 44}, 1023 (1986)]; D. Ebert,
R. N. Faustov and V. O. Galkin, Phys. Lett. B {\bf 537}, 241 (2002).
\vspace{-2.5mm}
\bibitem{fg} R. N. Faustov and V. O. Galkin, Z. Phys. C {\bf 66}, 119
(1995);  D. Ebert, R. N. Faustov and V. O. Galkin, Phys. Rev. D {\bf
  62}, 014032 (2000).
\vspace{-2.5mm}
\bibitem{mod} D. Ebert, R. N. Faustov and V. O. Galkin, Phys. Rev. D
  {\bf 56}, 312 (1997); R. N. Faustov, V. O. Galkin and
  A. Yu. Mishurov, Phys. Rev. D {\bf 53}, 6302 (1996); {\bf  53}, 1391
  (1996).
\vspace{-2.5mm}
\bibitem{efgbc}  D. Ebert, R. N. Faustov and V. O. Galkin, Phys. Rev. D
  {\bf 67}, 014027 (2003).
\vspace{-2.5mm}
\bibitem{3} A. A. Logunov and A. N. Tavkhelidze, Nuovo Cimento {\bf29},
380 (1963).
\vspace{-2.5mm}
\bibitem{4} A. P. Martynenko and R. N. Faustov, Theor. Math. Phys. {\bf
    64}, 765 (1985) [Teor. Mat. Fiz. {\bf 64}, 179 (1985)].
\vspace{-2.5mm}
\bibitem{ef} E. Eichten and F. Feinberg, Phys. Rev. D {\bf 23},
2724 (1981).
\vspace{-2.5mm}
\bibitem{schn} H. J. Schnitzer, Phys. Rev. D {\bf 18}, 3482 (1978).
\vspace{-2.5mm}
\bibitem{f} R. N. Faustov, Ann. Phys. {\bf 78}, 176 (1973); Nuovo
Cimento A {\bf 69}, 37 (1970).
\vspace{-2.5mm}
\bibitem{jlms} E. Jenkins, M. Luke, A.V. Manohar and M. Savage,
  Nucl. Phys. B {\bf 390}, 463 (1993).
\vspace{-2.5mm}
\bibitem{pdg}  Particle Data Group, K. Hagiwara {\it et al.},
  Phys. Rev. D {\bf 66}, 010001 (2002).
\vspace{-2.5mm}
\bibitem{wbc1} D. Ebert, R.N. Faustov and V.O. Galkin, Phys. Rev. D {\bf
  68}, 094020 (2003).
\vspace{-2.5mm}
\bibitem{wbc2} D. Ebert, R.N. Faustov and V.O. Galkin,
Eur. Phys. J. C {\bf 32}, 29 (2003).
\vspace{-2.5mm}
\bibitem{iks} M. A. Ivanov, J. G. K\"orner and P. Santorelli,
  Phys. Rev. D {\bf 63}, 074010 (2001).
\vspace{-2.5mm}
\bibitem{klo} V. V. Kiselev, A. K. Likhoded and A. I. Onishchenko,
  Nucl. Phys. B {\bf 569}, 473 (2000); V. V. Kiselev,
  {hep-ph/0211021}. 
\vspace{-2.5mm}
\bibitem{emv} A. Abd El-Hady, J. H. Mu\~noz and J. P. Vary, Phys. Rev. D
  {\bf 62}, 014019 (2000).
\vspace{-2.5mm}
\bibitem{cc} C.-H. Chang and Y.-Q. Chen, Phys. Rev. D {\bf 49}, 3399
  (1994). 
\vspace{-2.5mm}
\bibitem{cdf} P. Colangelo and F. De Fazio, Phys. Rev. D {\bf 61},
  034012 (2000).
\vspace{-2.5mm}
\bibitem{aknt} A. Yu. Anisimov, P. Yu. Kulikov, I. M. Narodetskii and
  K. A. Ter-Martirosyan, Phys. Atom. Nucl. {\bf 62}, 1739 (1999)
  [Yad. Fiz. {\bf 62}, 1868 (1999)].
\vspace{-2.5mm}
\bibitem{nw} M. A. Nobes and R. M. Woloshyn, J. Phys. G {\bf 26}, 1079
  (2000). 
\vspace{-2.5mm}
\bibitem{lyl} G. Lu, Y. Yang and H. Li, Phys. Lett. B {\bf 341}, 391
  (1995). 
\vspace{-2.5mm}
\bibitem{lc} J.-F. Liu and K.-T. Chao, Phys. Rev. D {\bf 56}, 4133
  (1997). 
\vspace{-2.5mm}
\bibitem{bsw} M. Bauer, B. Stech, and M. Wirbel, Z. Phys. C {\bf 34},
103 (1987).
\vspace{-2.5mm}
\bibitem{dg} M. J. Dugan and B. Grinstein, Phys. Lett. B {\bf 255}, 583
(1991).
\vspace{-2.5mm}
\bibitem{jb} J. D. Bjorken, Nucl. Phys. B (Proc. Suppl.) {\bf 11}, 325
(1989). 
\vspace{-2.5mm}
\bibitem{bbns} M. Beneke, G. Buchalla, M. Neubert and C. T. Sachrajda,
Phys. Rev. Lett. {\bf 83}, 1914 (1999); Nucl. Phys. B {\bf 591}, 313
(2000). 
\vspace{-2.5mm}
\bibitem{bs} A. J. Buras and L. Silvestrini, Nucl. Phys. B {\bf 569},
  3 (2000).
\end{thebibliography}
\end{document}